\documentclass[twocolumn,showpacs,preprintnumbers,amsmath,amssymb]{revtex4}

\usepackage{graphicx}
\usepackage{dcolumn}
\usepackage{bm}

\begin{document}

\title{A conjecture for turbulent flow}

\author{Zhao Jianglin
}
 \email{ydqc@eyou.com}
\affiliation{%
Institute of Mechanics (IMECH), Chinese Academy of Sciences,
Beijing, 100080, China
}%

\date{\today}

\begin{abstract}
In this paper, basing on a generalized Newtonian dynamics (GND)
approach which has been proposed elsewhere we present a conjecture
for turbulent flow. We firstly utilize the GND to reasonably unify
the two phenomenological methods recently proposed of the water
movement in unsaturated soils. Then in the same way a modified
Euler equation (MEE) is yielded. Under a zero-order approximation,
a simple split solution of the MEE can be obtained that shows flow
fluids would have a velocity field with the power-law scaling
feature (power-law fluid) for the case of high Reynolds number.
\end{abstract}


\maketitle

Turbulent flow is one of the  most bewildering  phenomena in
nature. Ever since a hundred years, people have initiated many
theories to explain it. These theories give different
phenomenological descriptions, and some of them are in good
agreement with experimental data and provide some very useful
clues for our understanding of true essence of turbulent flow.
However, up to now, we can not employ a unified framework to
describe turbulent flow yet, not to mention explaining it. we even
can not give a good definition of turbulence so that we can
quantitatively determine whether turbulent flow has appeared in a
fluid.

In this paper, we would present a conjecture for turbulent flow
with the generalized Newtonian dynamics (GND) in order to better
describe turbulence.

The GND describes the fractal world by means of  a fractional
dimensional kinetic velocity (or mass). In the anomalous
displacement variation model (ADVM), its one-dimensional basic
dynamics equation in terms of the Newtonian kinematics equation
form can be written as follows:
\begin{equation}
m\ddot{x}=\alpha(\sum_{i}F_{i})|x|^{1-q}+\kappa\frac{(\dot{x})^{2}}{x}.
\end{equation}
Here $m$ is mass of a particle, $F_{i}$ is the i-th external force
acting on the particle, $q$ is a velocity fractal index (vfi) and
$\dot x$ denotes $dx/dt$. The right-hand side of the generalized
Newtonian dynamics equation (1) can be understood as the effective
forces acting on the particle in the Euclidean space of the
fractal environment, and the two constants, $\alpha$ and $\kappa$,
which are given by $\alpha=m/(qm_{q})$ and $\kappa=m(1-q)$
respectively, are the effective force coefficients. Here $m_{q}$
is an effective mass of the particle in the fractal environment,
which has a dimension of $[kg\cdot m^{1-q}]$. The generalized
Newtonian dynamics Eq. (1) can be reduced to the Newtonian
kinematics equation $m\ddot{x}=\sum_{i}F_{i}$ when $q=1$ namely
the fractal velocity $\pm\frac{d|x|^{q}}{dt}$ becomes the ordinary
velocity $\frac{dx}{dt}$.

It is seen that the term $\kappa\frac{(\dot{x})^{2}}{x}$ in Eq.
(1) is an additional force which can be eliminated only when the
fractal environment around the particle disappears, namely $q=1$.
We call the term $\kappa\frac{(\dot{x})^{2}}{x}$ the fractal force
(FF). The FF appears in the complex fractal environment, its
additional property is similar to that of Coriolis force in a
rotational inertial system, but the origin or the practicable
situation of it is still unknown for us.

The cascade process in turbulence shows turbulent flow is a type
of the fractal phenomenon. Particularly scaling law discovered by
Kolmogorov \footnote{A.N.Kolmogorov. Compt Rend. Acod. Sci.
U.R.S.S. \textbf{30}, 301 (1941) ; \textbf{32}, 16 (1941).} and
the characteristic of the $\frac{1}{f}$ power spectrum of
turbulence signal \footnote{Liu Shi-Da and Liu Shi-Kuo.
\textit{Solitary Wave and Turbulence}. Shanghai Scientific and
Technological Education Publishing House, (1994).} can also
evidently exhibit that. Here we would assume all kinds of
turbulent flow as well as process from laminar flow or convection
to turbulence can be described on basis of the fractal geometry.
The fractal geometry, whose detail contents can be reviewed in
Ref. \footnote{B. B. Mandelbrot. \textit{The Fractal Geometry of
Nature}. W.H. Freeman and Company, (1988).}, was initialized by
Mandelbrot.

Accordingly, we want to introduce the GND into turbulence. We have
a definition of fractal velocity according to the anomalous
displacement variation model (ADVM):
\begin{equation}
v_{q}=\frac{dx'}{dt}=\pm\frac{d|x|^q}{dt}=q|x|^{q-1}\frac{dx}{dt}
\end{equation}
where $q$ is the velocity fractal index (vfi). We assume a
particle needs a $n\Delta t$ time length to ``jump'' a $(n\Delta
x)^{q}$ displacement length, then the instantaneous motion state
of the particle can be expressed by Eq. (2). Here $\Delta x$ and
$\Delta t$ is a fixed displacement length and time interval
respectively, and $n$ is a natural number representing the $n$-th
step of the particle. Because the instantaneous velocity of the
particle  can be written as
\begin{equation}
\ \ v_{n}=\lim_{\Delta t\to 0}\frac{(n\Delta x)^{q}-[(n-1)\Delta
x]^{q}}{\Delta t}
\end{equation}
\begin{equation}
=\lim_{\Delta t\to 0}\frac{(n\Delta x)^{q}-[(n-1)\Delta
x]^{q}}{\Delta x}\frac{\Delta x}{\Delta t} \ \ \ ;
\end{equation}
we let $\Delta x$=$\Delta|x(t)|$, $n\Delta x$=$|x(t)|$, and both
$\Delta|x(t)|$ and $\Delta t$ be an arbitrary small increment;
allowing for the directivity of velocity, Eq. (4) thus can  be
rewritten with $x=\pm|x|$ as Eq. (2) where $x'=\pm|x|^{q}$ which
represents the particle's displacement in the complex environment
(which is somewhat different from  real physical displacement).
Similarly, we have another form of definition of fractal velocity
according to the anomalous time variation model (ATVM):
\begin{equation}
v'_{q}=\frac{dx}{dt'}=\frac{dx}{dt^{q}}=\frac{1}{q}t^{1-q}\frac{dx}{dt}
\end{equation}
In Fig. 1, we give a schematic description of the two models.

\begin{figure}
\scalebox{0.6}{\includegraphics{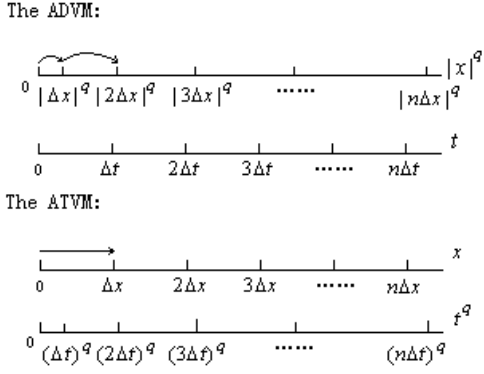}}\caption{\label{fig:epsart}
Schematic representation of the anomalous displacement variation
model (ADVM) and the anomalous time variation model (ATVM) of
anomalous motion. In the ADVM,  the particle takes the $n\Delta t$
time length to anomalously move (generally jump) the $(n\Delta
x)^{q}$ displacement length along the x-axis. In the ATVM,  the
particle needs the $(n\Delta t)^{q}$ time length to anomalously
walk (generally run and wait) the $n\Delta x$ displacement length
along the x-axis.}
\end{figure}

But, in fact, Eq. (2) and Eq. (5) are only different descriptions
of the same physical fact. For example, the fact that a particle
walks the $(n\Delta x)^q$ displacement length within the $n\Delta
t$ time interval can also be stated in another word that the
particle needs $(n\Delta t)^p$ time interval to walk $n\Delta x$
displacement length $(p\neq q)$. In addition, $\lambda^{1-q}x'$
and $t$ are real physical variables in Eq. (2) while $x$ is only a
scaling variable, where $\lambda$ denotes a characteristic length
and $m\lambda^{1-q}=m_{q}$; similarly $x$ and $\tau^{1-q}t'$ in
Eq. (5) are real physical variables while $t$ is only a scaling
variable, where $\tau$ is a characteristic time and
$m\tau^{2q-2}=m'_{q}$. Note that  $x-$space is changed to
$x^{q}-$space when a particle is moving with time, namely ordinary
scaling matching structure of time and space is modified, but any
force (except those related to velocity) is invariant from the
ordinary environment to the fractal space.

With the help of the GND approach, we can easily unify the two
phenomenological methods recently proposed of the water
percolation in unsaturated soils. In the unsaturated soil-water
transport, the water flux for the horizontal one-dimensional
column case follows the Buckingham-Darcy law:
\begin{equation}
Q=-k(\theta)\frac{\partial\psi}{\partial x}
\end{equation}
where $Q$ denotes the volume flow rate per unit area, $k$ is
unsaturated hydraulic conductivity, $\psi$ is the hydraulic pressure
head and $\theta$ is the soil water content; and the customary
Richards' equation
\begin{equation}
\frac{\partial \theta}{\partial t}=\frac{\partial}{\partial
x}[D(\theta)\frac{\partial\theta}{\partial x}]
\end{equation}
can be arrived at by combining Eq. (6) and the mass conservation
equation $\frac{\partial\theta}{\partial t}+\frac{\partial
Q}{\partial x}=0$ with $D(\theta)=k(\theta)\frac{d\psi}{d\theta}$.
Here $D(\theta)$ is the soil water diffusivity.

It is necessary to emphasize that movement of liquid water in
soils is not a diffusion phenomenon \footnote{D. Hillel, Soil and
Water: Physical Principles and Processes, 1971, Academic Press,
NY.}. Conversely, it is a kind of macroscopic flow motion which is
based on Navier-Stokes law.  Phillip derived the conclusion that
Buckingham-Darcy law follows from the Navier-Stokes equation with
$(\vec{u}\cdot\nabla)\vec{u}\simeq 0$:
$v_{k}\nabla^{2}\vec{u}=\nabla\phi$ where $v_{k}$ is dynamical
viscosity and $\phi$ is total potential \footnote{J.R, Philip,
Water Resources Res. \textbf{5}, 1070 (1969).}. Here no concept of
probability appears.

However, deviations from Eq. (7) \footnote{W. Gardner and J.A.
Widtsoe, Soil Sci. \textbf{11}, 215 (1921); Nielsen et al., Soil
Sci. Soc. Am. Proc. \textbf{26}, 107 (1962); S.I. Rawlins and W.H.
Gardner, Soil Sci. Soc. Am. Proc. \textbf{27}, 507 (1963); H.
Ferguson and W.R. Gardner, Soil Sci. Soc. Am. Proc. \textbf{27},
243 (1963).} cause further assumptions that the diffusivity has a
dependence on distance given by $|x|^{-\beta}D(\theta)$
\footnote{Y. Pachepsky and D. Timlin, J. Hydrology \textbf{204},
98 (1998).} or the diffusivity is a time-dependent quantity given
by $t^{2m-1}D(\theta)$ \footnote{I.A. Guerrini and D.
Swartzendruber, Soil Sci. Soc. Am. J. \textbf{56}, 335 (1992).}.
Here $m$ is a positive constant. We now want to unify the above
two assumptions and do not need to introduce other models for
microscopic anomalous transport.

In the GND framework, the reduced Navier-Stokes equation is
rewritten as $v'_{k}\nabla^{2}\vec{u}_{s}=\nabla\phi$ where
$\vec{u}_{s}$ is fractal velocity.  Then the Richards' equation is
generalized as
\begin{equation}
\frac{\partial\theta}{\partial
t}=\frac{1}{s}\frac{\partial}{\partial
x}(|x|^{1-s}D_{s}(\theta)\frac{\partial\theta}{\partial x})
\end{equation}
and
\begin{equation}
\frac{\partial\theta}{\partial t}=s\frac{\partial}{\partial
x}(t^{s-1}D_{s}(\theta)\frac{\partial\theta}{\partial x})\ ,
\end{equation}
respectively. Here $s$ is the vfi and $D_{s}(\theta)$ is a fractal
soil water diffusivity.

Eq. (8) and Eq. (9) are similar to the generalized Richards'
equations proposed in [9] and [10], respectively, which have been
shown to be valid for some experiment data. But we can unify the
two kinds of assumption in one framework, and it seems more
reasonable because previous frameworks of anomalous diffusion are
not appropriated for describing essentially macroscopic flow
movements.

\begin{figure}
\scalebox{0.6}{\includegraphics{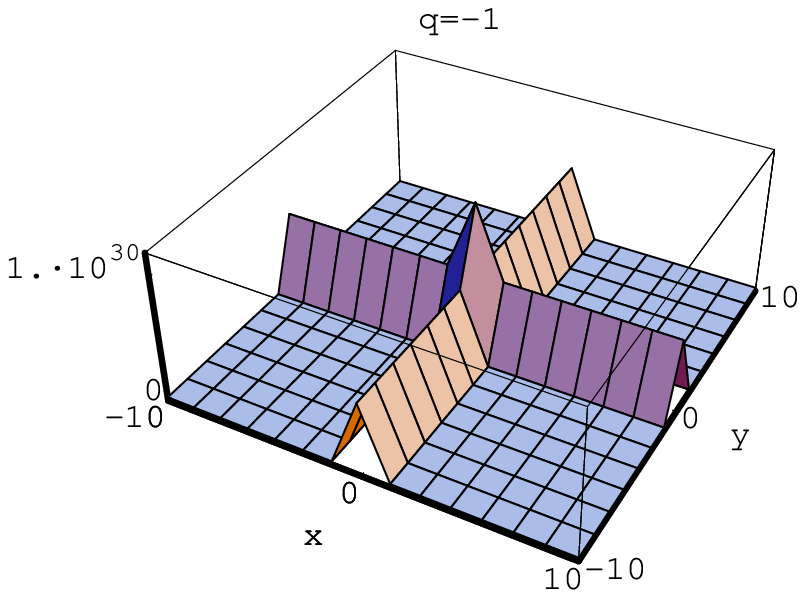}}
\end{figure}
\begin{figure}
\scalebox{0.6}{\includegraphics{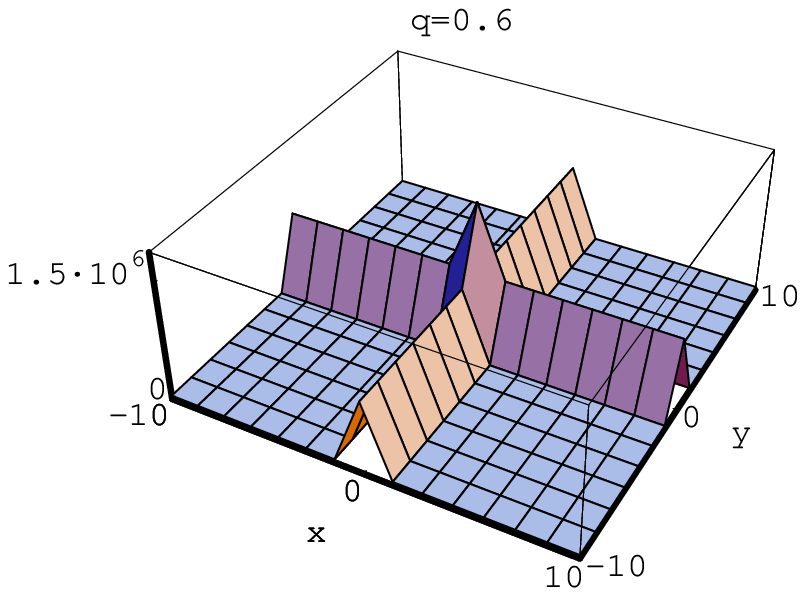}}
\end{figure}
\begin{figure}
\scalebox{0.6}{\includegraphics{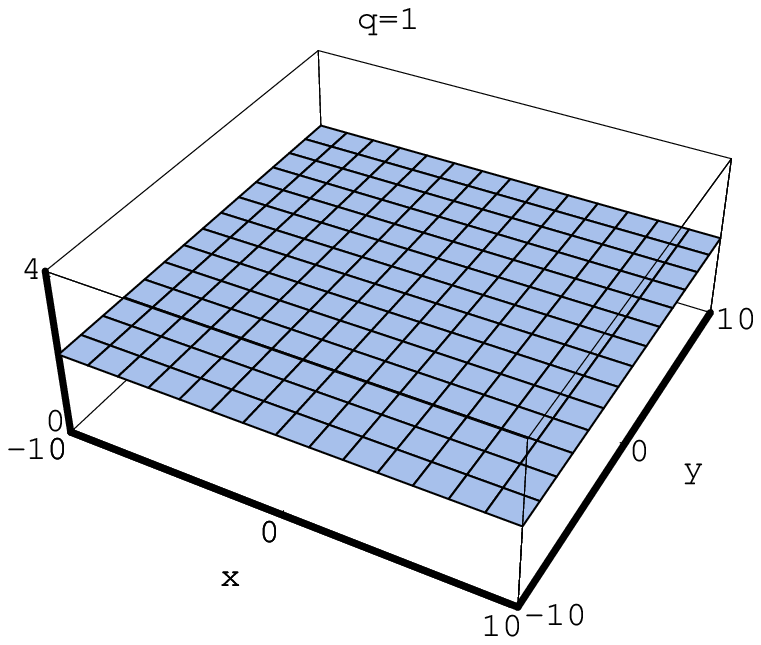}}
\end{figure}
\begin{figure}
\scalebox{0.6}{\includegraphics{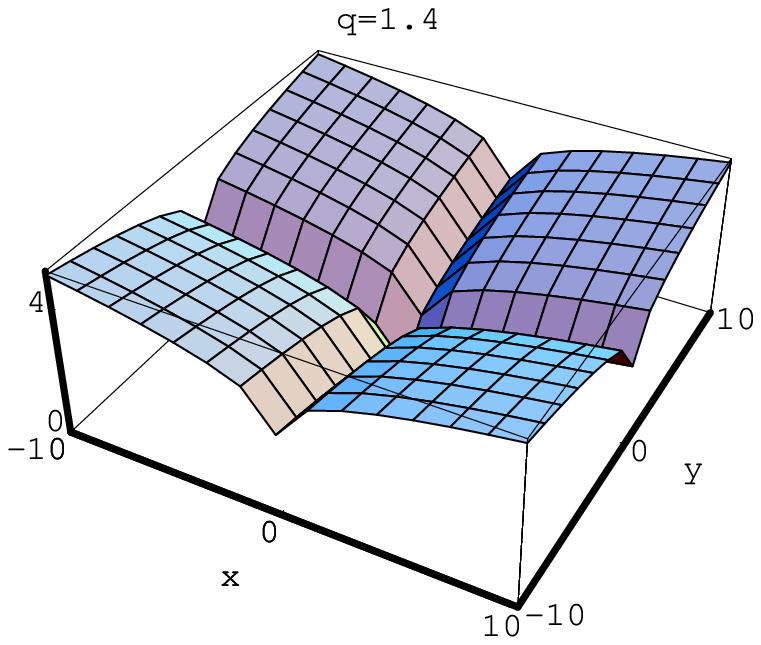}}
\end{figure}
\begin{figure}
\scalebox{0.6}{\includegraphics{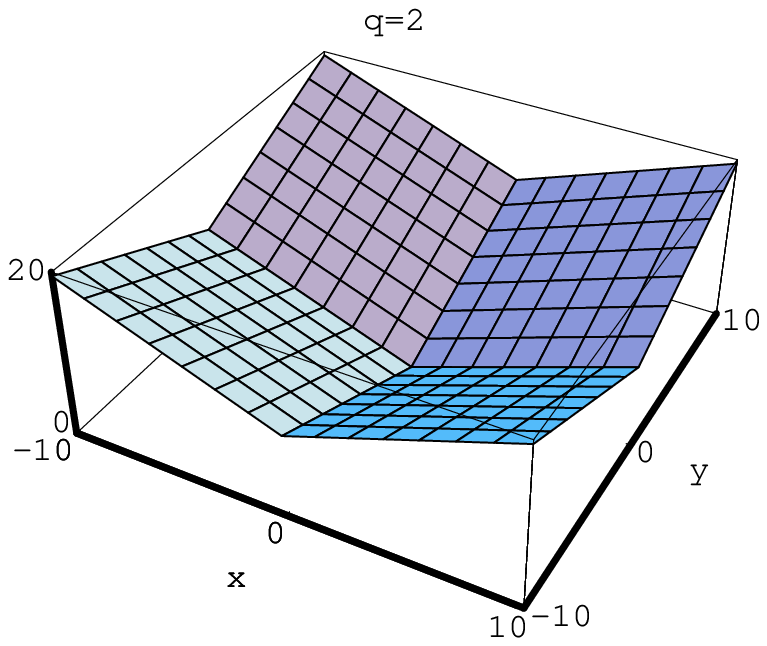}}
\end{figure}
\begin{figure}
\scalebox{0.6}{\includegraphics{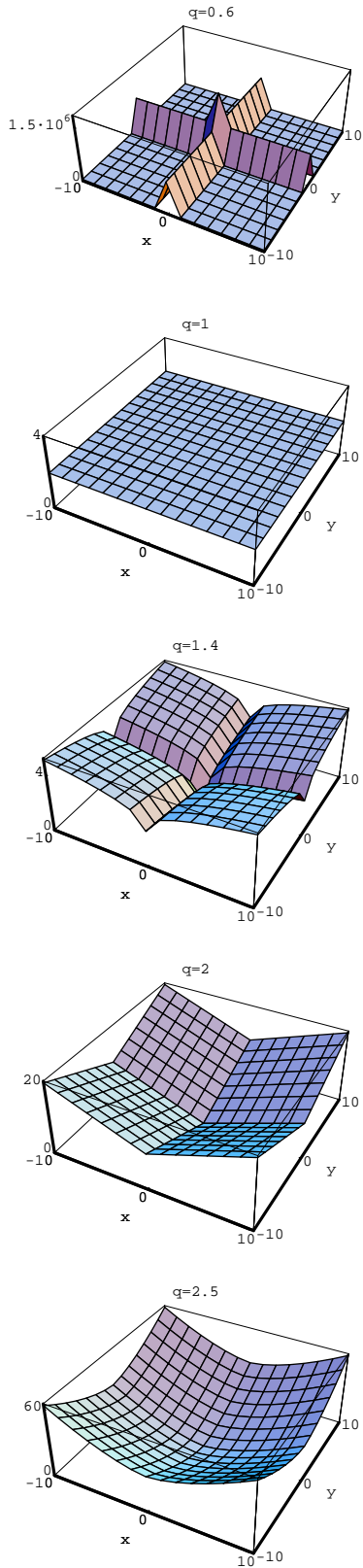}}\caption{\label{fig:epsart}
Velocity field $v_{x}(x,y)$ given by a solution of the MEE Eq.
(19) for different values of $q$. The solution reflects motion
state of a flow in the fractal environment. Here $a$ and $b$ are
both taken 1.}
\end{figure}

For the turbulence problem, we can obtain a modified Euler
equation (MEE) from the GNKE Eq. (1) according to the principle of
the GND:
\begin{equation}
\frac{\partial \vec{v}}{\partial t}+(\vec{v}\cdot
\nabla)\vec{v}=-\frac{\alpha}{\rho}\nabla_{q}p+\frac{\alpha}{\rho}\vec{F}_{q}+\beta\vec{a}_{f}
\end{equation}
where $\alpha=\frac{m}{qm_{q}}$, $\beta=1-q$,

\begin{equation}
\nabla_{q}=\begin{pmatrix} |x|^{1-q}\frac{\partial}{\partial x}
\\ |y|^{1-q}\frac{\partial}{\partial y} \\
|z|^{1-q}\frac{\partial}{\partial z} \end{pmatrix} \ \ ,
\end{equation}

\begin{equation}
\vec{F_{q}}=\begin{pmatrix} |x|^{1-q}\frac{dF_{x}}{dV}
\\ |y|^{1-q}\frac{dF_{y}}{dV} \\
|z|^{1-q}\frac{dF_{z}}{dV} \end{pmatrix}\ \ ,
\end{equation}

\begin{equation}
\vec{a_{f}}=\begin{pmatrix} \frac{(\dot{x})^{2}}{x} \\
\frac{(\dot{y})^{2}}{y} \\
\frac{(\dot{z})^{2}}{z} \end{pmatrix}\ \ ,
\end{equation}
\\$\rho$ is density of a fluid; $-\alpha dV\nabla_{q} p$ is an effective pressure acting
on certain a volume element $dV$ from surrounding fluid bodies and
$\vec{F}_{q}$ is considered as an effective complex force field
acting on a volume element $dV$ of the fluid where $dF_{i}$
$(i=x,y,z)$ is a real physical force (other than pressure) acting
on a volume element $dV$. We notice that $\vec{a}_{f}$ has the
same dimension as another nonlinear term
$(\vec{v}\cdot\nabla)\vec{v}$.

We call $\nabla_{q}$  the fractional gradient operator, and also
we notice that $\rho\beta\vec{a_{f}}$ is exactly the fractal force
on a unit volume of fluid. In addition, the coefficient $\alpha$
with the dimension [$m^{q-1}$] can be seen as $\lambda^{q-1}$.
Here $\lambda$ is a characteristic length of the fluid. We see the
MEE (10) will will recover the ordinary Euler equation:
\begin{equation}
\frac{\partial \vec{v}}{\partial t}+(\vec{v}\cdot
\nabla)\vec{v}=-\frac{1}{\rho}\nabla
p+\frac{1}{\rho}\frac{d\vec{F}}{dV} \ .
\end{equation}
Here $d\vec{F}$ is  force (other than pressure) acting on the
volume element $dV$.

The MEE reflects the situation of a fluid with fractal feature,
since it is derived directly from a fractal velocity which
describes the fractal state of a system. We think the ordinary
Euler equation can not play a role in determining the fractal
state of the fluid, for the nonlinear term $(\vec{v}\cdot
\nabla)\vec{v}$ is only a kinematic factor rather than a dynamical
factor like the term $(1-q)\vec{a}_{f}$ in the MEE. In other
words, when the fluid becomes in a fractal state, the fluid would
be acted on by a special fractal force which appears only in the
fractal situation, and the fractal force can also be seen as an
external energy input; at the same time, ordinary forces will also
be deformed, but these changes can not follow naturally from the
ordinary Euler equation. Of course, the process from a laminar
flow or a convection current to turbulent flow is controlled by
real physical parameters such as velocity of the fluid or
temperature difference, these parameters often can be combined
into a dimensionless number such as the Reynolds number $Re$.
Particularly there is a critical Reynolds number $Rec$. When the
Reynolds number of a fluid is above $Rec$, turbulence happens.
That can make us guess that the vfi $q$ is a function of these
dimensionless numbers and critical dimensionless numbers such as
$Re$ and $Rec$ such that the fact that turbulence emerges when
$Re>Rec$ is equal to the situation of the MEE for $q>1$ (or
$q<1$).

The viscous term of a fluid can be directly added in $\vec{F_{q}}$
in Eq. (10), but here we would take a look at solutions of the MEE
in a range of high Reynolds number.

We study the MEE for the case of the two-dimensional
incompressible and constant flow, and we consider $\vec{F}_{q}=0$;
then Eq. (10) becomes
\begin{equation}
\left\{ \begin{array}{c} v_{x}|x|^{q-1}\frac{\partial
v_{x}}{\partial x}+v_{y}|x|^{q-1}\frac{\partial v_{x}}{\partial
y}=-\frac{\alpha}{\rho}\frac{\partial p}{\partial
x}+\beta|x|^{q-1}\frac{v^{2}_{x}}{x} \\
v_{x}|y|^{q-1}\frac{\partial v_{y}}{\partial
x}+v_{y}|y|^{q-1}\frac{\partial v_{y}}{\partial
y}=-\frac{\alpha}{\rho}\frac{\partial p}{\partial
y}+\beta|y|^{q-1}\frac{v^{2}_{y}}{y}
\end{array}\right. \
\end{equation}

We make such an approximation that
\begin{equation}
\left\{ \begin{array}{c} a|x|^{q-1}\frac{\partial v_{x}}{\partial
x}+b|x|^{q-1}\frac{\partial v_{x}}{\partial
y}=-\frac{\alpha}{\rho}\frac{\partial p}{\partial
x}+\beta|x|^{q-1}\frac{v^{2}_{x}}{x} \\
a|y|^{q-1}\frac{\partial v_{y}}{\partial
x}+b|y|^{q-1}\frac{\partial v_{y}}{\partial
y}=-\frac{\alpha}{\rho}\frac{\partial p}{\partial
y}+\beta|y|^{q-1}\frac{v^{2}_{y}}{y}
\end{array}\right. \
\end{equation}
where $a$ and $b$, which represent the mean velocity of the fluid
$\bar{v}_{x}$ and $\bar{v}_{y}$, respectively, are constant. The
approximation is reasonable  for a fluid with  small velocity
gradient.

Eliminating the pressure intensity $p$ and utilizing the
continuity condition,
\begin{equation}
\frac{\partial v_{x}}{\partial x}+\frac{\partial v_{y}}{\partial
y}=0\ ,
\end{equation}
we have
\begin{equation}
|x|^{q-1}\frac{\partial^{2}\psi}{\partial
y^{2}}+|y|^{q-1}\frac{\partial^{2}\psi}{\partial
x^{2}}=\beta\frac{|x|^{q-1}}{x}\frac{\partial v^{2}_{x}}{\partial
y}-\beta\frac{|y|^{q-1}}{y}\frac{\partial v^{2}_{y}}{\partial x}
\end{equation}
where $\psi=bv_{x}-av_{y}$.

We let $v^{2}_{x}=\frac{(\psi+av_{y})^{2}}{b^2}\sim \psi^{2}$ as
well as $v^{2}_{y}\sim \psi^{2}$, which is proper for a fluid
whose velocity is not too large. Then we obtain a second-order
partial differential equation concerning $\psi$:
\begin{equation}
|x|^{q-1}\frac{\partial^{2}\psi}{\partial
y^{2}}+|y|^{q-1}\frac{\partial^{2}\psi}{\partial
x^{2}}=\frac{\beta}{b^{2}}\frac{|x|^{q-1}}{x}\frac{\partial\psi^{2}}{\partial
y}-\frac{\beta}{a^{2}}\frac{|y|^{q-1}}{y}\frac{\partial\psi^{2}}{\partial
x}\ .
\end{equation}

We neglect the two second partial derivative terms of $\psi$,
because both of them  can be seen as small quantities. Eq. (19)
thus is simplified as a first-order partial differential equation:
\begin{equation}
b^{2}x|x|^{1-q}\frac{\partial\psi}{\partial
x}=a^{2}y|y|^{1-q}\frac{\partial\psi}{\partial y}
\end{equation}
whose general solution can be easily arrived at:
\begin{equation}
\psi=\Phi(u)=\Phi(a^2|x|^{q-1}+b^{2}|y|^{q-1})\ .
\end{equation}
Here $\Phi$ is any continuous and differentiable function of $u$.

For simplicity, we choose $\psi$ as follows:
\begin{equation}
bv_{x}-av_{y}=(b-a)(a^{2}|x|^{q-1}+b^{2}|y|^{q-1})\ ,
\end{equation}
such that
\begin{equation}
v_{x}=v_{y}=a^{2}|x|^{q-1}+b^{2}|y|^{q-1}\ .
\end{equation}
Evidently, the velocity field $v_{x}$ or $v_{y}$ is a parabolic
dish  for different values of $q$.

Therefore, under such an approximation, we obtain a simple
solution of the MEE for a two-dimensional incompressible and
constant flow in the high Reynolds number limit:
\begin{equation}
v_{i,j}=\bar{v}_{i}^{2}|x_{i}|^{q-1}+\bar{v}_{j}^{2}|x_{j}|^{q-1}
\end{equation}
where $v_{i,j}$ denotes $v_{i}$ or $v_{j}$.

We graph velocity field $v_{x}$ given by Eq. (23) for different
values of $q$ in Fig. 2. Here $a$ and $b$ are both taken 1. It is
clearly seen that manifold is broken when $q<1$ and the fluid
velocity $v_{x}$ becomes infinity at the position $x=0$ or $y=0$.
We can imagine that the fluid body could be broken up into four
closed packets in each dimension if proper boundary conditions
about ($x, y, v_{i,j}$) are given.  Thus we can call the solution
Eq. (23) for the case of $q<1$ a split solution for the MEE. The
situation for $1<q<2$ is somewhat different where four convexes of
the fluid are not broken.  When $q>2$, velocity field of the fluid
becomes an concave parabolic dish. For $q=2$ the velocity field
 comprises four planes and for $q=1$ a laminar flow case is
recovered.

The solution reflects motion state of a flow in the fractal
environment which is embodied in the MEE by both a fractal force
which can be understood as an external nonlinear energy input and
a fractional gradient operator $\nabla_{q}$ which is similar to
the Riesz fractional derivative operator \footnote{K. B. Oldham
and J. Spanier, \textit{The Fractional Calculus} (Academic Press,
New York, 1974).}\footnote{S. G. Samko, A. A. Kilbas, and O. I.
Marichev, \textit{Fractional Integrals and Derivatives: Theory and
Applications} (Gordon and Breach, New York, 1993).}. It is
important that the fractional gradient operator will
simultaneously be reduced to the ordinary gradient operator when
the fractal force disappears. However, whether the MEE can
describe turbulent flow or not is still an open problem.

Summarizing, the fact that the GND framework can reasonably unify
the two phenomenological methods recently proposed of anomalous
transport of water in unsaturated soils and other satisfactory
results of the GND elsewhere make us conjecture that the GND
constitutes a dynamical basis of the problem of turbulence, thus a
modified Euler equation (MEE) is yielded. Under a zero-order
approximation, a simple split solution of the MEE can be obtained,
and we see that turbulent flow would have  a velocity field with
the power-law scaling feature for the case of high Reynolds
number.

\end{document}